\documentclass[aps,pre,final,twocolumn]{revtex4}

\usepackage{amsmath,bbm,graphicx,bm}

\DeclareMathOperator{\re}{Re}
\DeclareMathOperator{\im}{Im}
\DeclareMathOperator{\sn}{sn}
\DeclareMathOperator{\dn}{dn}
\DeclareMathOperator{\cn}{cn}
\DeclareMathOperator{\sgn}{sgn}
\newcommand{\deriv}[2]{\frac{\partial #1}{\partial #2}}

\begin{document}

\title{Discontinuous Molecular Dynamics for Semi-Flexible and Rigid Bodies}

\author{Lisandro Hern\'{a}ndez de la Pe\~{n}a, Ramses van Zon, 
        Jeremy Schofield} 

\affiliation{Chemical Physics Theory Group, Department of Chemistry,
             University of Toronto, Ontario, Canada M5S 3H6}

\author{Sheldon B. Opps} 

\affiliation{Department of Physics, University of Prince Edward Island,
         550 University Avenue, Charlottetown, Prince Edwards Island,
         Canada C1A 4P3}

\date{July 20, 2006}

\begin{abstract}
A general framework for performing event\mbox-driven simulations of
systems with semi\mbox-flexible or rigid bodies interacting under
impulsive torques and forces is outlined. Two different approaches are
presented.  In the first, the dynamics and interaction rules are
derived from Lagrangian mechanics in the presence of constraints.
This approach is most suitable when the body is composed of relatively
few point masses or is semi\mbox-flexible.  In the second method, the
equations of rigid bodies are used to derive explicit analytical
expressions for the free evolution of arbitrary rigid molecules and to
construct a simple scheme for computing interaction rules.  Efficient
algorithms for the search for the times of interaction events are
designed in this context, and the handling of missed interaction
events is discussed.
\end{abstract}

\maketitle

\section{Introduction}

There has been an increasing interest over the last decade in
performing large\mbox-scale simulations of colloidal systems, proteins,
micelles and other biological assemblies. Simulating such systems, and
the phenomena that take place in them, typically requires a
description of dynamical events that occur over a wide range of time
scales. Nearly all simulations of such systems to date are based on
following the microscopic time evolution of the system by integration
of the classical equations of motion. Usually, due to the complexity
of intermolecular interactions, this integration is carried out in a
step\mbox-by\mbox-step numerical fashion producing a time ordered set of
phase\mbox-space points (a \emph{trajectory}).  This information can then
be used to calculate thermodynamic properties, structural functions or
transport coefficients.  An alternative approach, which has been
employed in many contexts, is to use step potentials to approximate
intermolecular interactions while affording the analytical solution of
the dynamics
\cite{AlderWainwright59,AlderWainwright60,AllenTildesley,Allenetal89,Rapaport}.
The simplification in the interaction potential can lead to an
increase in simulation efficiency since the demanding task of
calculating forces is reduced to computing momentum exchanges between
bodies at the instant of interaction. This approach is called
event\mbox-driven or \emph{discontinuous molecular dynamics} (DMD).
 
In the DMD approach, various components of the system interact via
discontinuous forces, leading to impulsive forces that act at specific
moments of time.  As a result, the motion of particles is free of
inter\mbox-molecular forces between impulsive \emph{events} that alter the
trajectory of bodies via discontinuous jumps in the momenta of the
system at discrete interaction times.  To determine the dynamics, the
basic interaction rules of how the (linear and angular) momenta of the
body are modified by collisions must be specified.

For molecular systems with internal degrees of freedom it is
straightforward to design fully\mbox-flexible models with discontinuous
potentials, but DMD simulations of such systems are often inefficient
due to the relatively high frequency of internal
motions\cite{Chapelaetal89}.  This inefficiency is reflected by the
fact that most collision events executed in a DMD simulation
correspond to intra rather than inter\mbox-molecular interactions.  On the
other hand, much of the physics relevant in large\mbox-scale simulations
is insensitive to details of intra\mbox-molecular motion at long times.
For this reason, methods of incorporating constraints into the
dynamics of systems with continuous potentials have been developed
that eliminate high frequency internal motion, and thus extend the
time scales accessible to simulation.  Surprisingly, relatively little
work has appeared in the literature on incorporating such constraints
into DMD simulations.  The goal of this paper is to extend the
applicability of DMD methods to include constrained systems and to
outline efficient methods that are generally applicable in the
simulations of semi\mbox-flexible and rigid bodies interacting via
discontinuous potentials.

In contrast to systems containing only simple spherical particles
\cite{Allenetal89,Rapaport,Chapelaetal84,Chapelaetal89,VanZonSchofield02b},
the application of DMD methods to rigid\mbox-body systems is complicated
by two main challenges.  The first challenge is to analytically solve
the dynamics of the system so that the position, velocity, or angular
velocity of any part of the system can be obtained exactly.  This is
in principle possible for a rigid body moving in the absence of forces
and torques, even if it does not possess an axis of symmetry which
facilitates its motion.  However, an explicit solution suitable for
numerical implementation seems to be missing in the literature
(although partial answers are abundant
\cite{Rueb1834,Jacobi1850,Whittaker,LandauLifshitzM,MarsdenRatiu,Masutanietal94}).
For this reason, we will present the explicit solution here.  Armed
with a solution of the dynamics of all bodies in the system, one can
calculate the collision times in an efficient manner, and in some
instances, analytically.

The second challenge is to determine how the impulsive forces lead to
discontinuous jumps in the momenta of the interacting bodies.  For
complicated rigid or semi\mbox-flexible bodies, the rules for computing
the momentum jumps are not immediately obvious.  It is clear however
that these jumps in momenta must be consistent with basic conservation
laws connected to symmetries of the underlying Lagrangian
characterizing the dynamics.  Often the basic Lagrangian is invariant
to time and space translations, and rotations, and, hence, the rules
governing collisions must explicitly obey energy, momentum, and
angular momentum constraints.  Such conservation laws can be utilized
as a guide to derive the proper collision rules.

A first attempt to introduce constraints into an event\mbox-driven system
was carried out by Ciccotti and Kalibaeva\cite{CiccottiDiatomic}, who
studied a system of rigid, diatomic molecules (mimicking liquid
nitrogen). Furthermore, non\mbox-spherical bodies of a special kind were
treated by Donev {\it et al.}\cite{Donevetal05a,Donevetal05b} by
assuming that all rotational motion in between interaction events was
that of a spherically symmetric body.  More recently, a spherically
symmetric hard\mbox-sphere model with four tetrahedral short ranged
(sticky) interactions (mimicking water) has been studied by De Michele
{\it et al.}\cite{DeMichele} with an event\mbox-driven molecular dynamics
simulation method similar to the most basic scheme presented in this
paper.  This work primarily focuses on the phase diagram of this
``sticky'' water model as a prototype of network forming molecular
systems.  Our purpose, in contrast, is to discuss a general framework
that allows one to carry out event\mbox-driven DMD simulations in the
presence of constraints and, in particular, for fully general rigid
bodies.  The methodology is applicable to modeling the correct
dynamics of water molecules in aqueous solutions\cite{water} as well
as other many body systems.

The paper is organized as follows.  Section~\ref{Calculation}
discusses the equations of motions in the presence of constraints and
Sec.~\ref{IncludingCollisions} discusses the calculation and
scheduling of collision times. The collision rules are derived in
Sec.~\ref{Rules}. In Sec.~\ref{Ensembles} it is shown how to sample
the canonical and microcanonical ensembles and how to handle subtle
issues concerning missing events that are particular to event\mbox-driven
simulations.  Finally, conclusions are presented in
Sec.~\ref{Conclusions}.

\section{Equations of Motion with Constraints}
\label{Calculation}

\subsection{Constrained dynamics}
\label{ConstrainedDynamics}

The motion of rigid bodies can be
considered to be a special case of the dynamics of systems under a
minimal set of $c$ time-independent holonomic constraints (i.e.\
dependent only on positions) that fix all intra\mbox-body distances:
\begin{equation}
  \sigma_\alpha (\bm r_N) = 0,
\end{equation}
where the index $\alpha$ runs over all constraints present in the
system and $\bm r_N$ is a generalized vector whose components are the
set of all Cartesian coordinates of the $N$ total particles in the
system. For fully rigid bodies, the number of constraints $c$ can
easily be calculated by noting that the number of spatial degrees of
freedom of an $n$\mbox-particle body is $3n$ in 3 dimensions, while only 6
degrees of freedom are necessary to completely specify the position of
all components of a rigid body: 3 degrees of freedom for the center of
mass of the object and 3 degrees of freedom to specify its orientation
with respect to some arbitrary fixed reference frame.  There are
therefore $c=3n-6$ constraint equations for a single rigid body with
$n$ point masses.  Below, these point masses will be referred to as
\emph{atoms} while the body as a whole will be called a
\emph{molecule}.  A typical constraint equation fixes the distance
between atoms $i$ and $j$ in the molecule to be some value, $d$, i.e.,
\begin{equation}
  \sigma (r_{ij}) = \frac{1}{2} \left( r_{ij}^{2} - d^2 \right) = 0.
\end{equation}
The equations of motion for the system follow from Hamilton's
principle of stationary action, which states that the motion of the
system over a fixed time interval is such that the variation of the
line integral defining the action $\mathcal S$ is zero:
\begin{eqnarray}
  \mathcal S &=& \int_{t_1}^{t_2} \!\mathcal L 
  (\bm r_N(t),\dot{\bm r}_N(t))\,dt 
\\
  \delta \mathcal S &=& \delta \int_{t_1}^{t_2} \!
  \mathcal L(\bm r_N(t),\dot{\bm r}_N(t)) \, dt = 0,
\end{eqnarray}
where the Lagrangian $\mathcal L$ in the presence of the constraints
is written in Cartesian coordinates as
\begin{equation}
  \mathcal L (\bm r_N,\dot{\bm r}_N) = 
  \sum_{i=1}^N\frac{m_i}{2} |\dot{\bm r}_i|^2 -
  \lambda_\alpha \sigma_\alpha (\bm r_N) - \Phi (\bm r_N),
\label{Lagrangian}
\end{equation}
where $\Phi$ is the interaction potential. For clarity throughout this
paper, the Einstein summation convention will be used for sums over
repeated \emph{greek} indices i.e.\ $\lambda_\alpha \sigma_\alpha
\equiv \sum_{\alpha=1}^c \lambda_\alpha \sigma_\alpha$, whereas the
sum over atom indices will be written explicitly.  In
Eq.~\eqref{Lagrangian}, the parameters $\lambda_\alpha$ are Lagrange
multipliers to enforce the distance constraints $\sigma_\alpha$.  The
resulting equations of motion are:
\begin{eqnarray}
  \frac{d}{dt} \deriv{\mathcal L}{\dot{\bm r}_i} &=& 
  \deriv{\mathcal L}{\bm r_i}
\\
  m_i \ddot{\bm r}_i &=& 
  - \lambda_\alpha \deriv{\sigma_\alpha}{\bm r_i} - \deriv{\Phi}{\bm r_i}.
\label{ELeq} 
\end{eqnarray}
For an elementary discussion of constrained dynamics in the Lagrangian
formulation of mechanics, we refer to Ref.~\onlinecite{Goldstein}.

When there are no interactions, such as for a single molecule, the
potential $\Phi=0$ and Eq.~\eqref{ELeq} becomes
\begin{equation}
  m_i \ddot{\bm r}_i = - \lambda_\alpha \deriv{\sigma_\alpha}{\bm r_i}.
\label{eom}
\end{equation}
These equations of motion must be supplemented by equations for the
$c$ Lagrange multipliers $\lambda_\alpha$, which are functions of
time. Although the $\lambda_\alpha$ are not functions of $\bm r_N$ and
$\dot{\bm r}_N$ in a mathematical sense, it will be shown below that
once the equations are solved they can be expressed in terms of $\bm
r_N$ and $\dot{\bm r}_N$.  Note that the equations of motion show that
even in the absence of an external potential, the motion of the point
masses (atoms) making up a rigid body (molecule) are non\mbox-trivial due
to the emergence of a \emph{constraint force} $- \lambda_\alpha
\partial \sigma_\alpha / \partial \bm r_i$.

In fortuitous cases, the time dependence of the Lagrange multipliers
is relatively simple and can be solved for by Taylor expansion of the
Lagrange multipliers in time $t$.  To evaluate the time derivatives of
the multipliers, one can use time derivatives of the initial
constraint conditions, which must vanish to all orders.  The result is
a hierarchy of equations, which, at order $k$, is linear in the
unknown $k$th time derivatives $\lambda_\alpha^{(k)}$ but depends on
the lower order time derivatives $\lambda_\alpha^{(0)}$,
$\lambda_\alpha^{(1)}$, \dots $\lambda_\alpha^{(k-1)}$.  In
exceptional circumstances, this hierarchy naturally truncates.  For
example, for a rigid diatomic molecule with a single bond length
constraint, one finds that the hierarchy truncates at order $k=0$, and
the Lagrange multiplier is a constant \cite{CiccottiDiatomic}. However
this is not the typical case.

Alternatively, since the constraints $\sigma_\alpha(\bm r_N)=0$ are to
be satisfied at all times $t$, and not just at time zero, their time
derivatives are zero at all times. {}From the first time derivative
\[
  \dot{\sigma}_\alpha(\bm r_N) = 0,
\]
one sees that the initial velocities $\bm v_i = \dot{\bm r}_i$ must
obey
\begin{equation}
\sum_i \bm v_i \cdot \deriv{\sigma_\alpha}{\bm r_i} = 0,
\label{mustobey}
\end{equation}
for each constraint condition $\alpha$.  The Lagrange multipliers can
be determined by the condition that the second derivatives of all the
constraints vanish so that
\begin{eqnarray*}
  \ddot{\sigma}_\alpha (\bm r_N) 
  &=& \sum_i \deriv{\sigma_\alpha
    (\bm r_N)}{\bm r_i} \cdot \ddot{\bm r}_i + \sum_{i,j} \dot{\bm r}_j \cdot
  \frac{\partial^{2} \sigma_\alpha(\bm r_N)}{\partial \bm r_j
    \partial\bm r_i} \cdot \dot{\bm r}_i  
\\
  &=& - \sum_i \deriv{\sigma_\alpha
    (\bm r_N)}{\bm r_i} \cdot
  \frac{\lambda_\beta}{m_i} \deriv{\sigma_\beta (\bm r_N)}{\bm r_i} 
\\
  && +
  \sum_{i,j} \dot{\bm r}_j \cdot
  \frac{\partial^{2} \sigma_\alpha(\bm r_N)}{\partial \bm r_j
    \partial\bm r_i} \cdot \dot{\bm r}_i
\\
  &=&0,
\end{eqnarray*}
yielding a linear equation for the Lagrange multipliers that can be
solved in matrix form as
\begin{equation}
  \lambda_\alpha(t) 
  =
  \mathbf Z_{\alpha\beta}^{-1}(\bm r_N(t)) 
  \mathcal{T}_\beta(\bm r_N(t),\dot{\bm r}_N(t)),
\label{DynamicMultipliers}
\end{equation}
where
\begin{eqnarray}
  \mathcal{T}_\beta(\bm r_N,\dot{\bm r}_N) 
  &=& \sum_{i,j}\dot{\bm r}_j \cdot
  \frac{\partial^{2} \sigma_\beta(\bm r_N)}{\partial \bm r_j
    \partial\bm r_i} \cdot \dot{\bm r}_i \label{Tdef} 
\\
  \mathbf Z_{\alpha \beta}(\bm r_N) 
  &=& \sum_i \frac{1}{m_i} \deriv{\sigma_\alpha
    (\bm r_N)}{\bm r_i} \cdot 
  \deriv{\sigma_\beta (\bm r_N)}{\bm r_i} .
\label{Zdef}
\end{eqnarray}
It may be shown that with $\lambda_\alpha$ given by
Eq.~\eqref{DynamicMultipliers}, all higher time derivatives of
$\sigma_\alpha$ are automatically zero.

As Eq.~\eqref{DynamicMultipliers} shows, in general the Lagrange
multipliers are dependent on both the positions $\bm r_N$ and the
velocities $\dot{\bm r}_N$ of the particles. To see that this makes
the dynamics non\mbox-Hamiltonian, the equations of motion can be cast
into Hamiltonian\mbox-like form using $\bm p_i=m_i\dot{\bm r}_i$, i.e.,
\begin{equation}
\label{nheom}
\begin{split}
  \dot{\bm r}_i &= \frac{\bm p_i}{m_i} \\
  \dot{\bm p}_i &= - \lambda_\alpha\deriv{\sigma_\alpha}{\bm r_i},
\end{split}
\end{equation}
where it is apparent that the forces in the system depend on the
momentum through $\lambda$ in Eq.~\eqref{DynamicMultipliers}.  There
exists no Hamiltonian that generates these equations of
motion.\cite{fna}

Since the underlying dynamics of the system is non\mbox-Hamiltonian, the
statistical mechanics of the constrained system is potentially more
complex.  In general, phase\mbox-space averages have to be defined with
respect to a metric that is invariant to the standard measure of
Hamiltonian systems, but $d\bm r_Nd\bm p_N$ is not conserved under the
dynamics and the standard form of the Liouville equation does not hold
\cite{Ramshaw,Tuckerman1}.  In general, there is a phase\mbox-space
compressibility factor $\kappa$ associated with the lack of
conservation of the measure that is given by minus the divergence of
the flow in phase space. It may be shown
that\cite{Tuckerman1,Tuckerman2}
\[
\kappa =
-\deriv{}{\bm r_i} \cdot \dot{\bm r}_i -\deriv{}{\bm p_i} \cdot \dot{\bm p}_i 
= \frac{d}{dt} \ln \| \mathbf Z(\bm r_N) \| ,
\]
where $\| \mathbf Z (\bm r_N) \|$ is the determinant of the matrix
$\mathbf Z_{\alpha\beta}(\bm r_N)$ defined in Eq.~(\ref{Zdef}).  The
compressibility factor is related to the invariant phase\mbox-space metric
$d\mu = \sqrt{g(\bm r_N,\bm p_N,t)} d\bm r_N d\bm p_N$
with\cite{Tuckerman2,Melchionna}
\begin{equation}
  \sqrt{g(\bm r_N,\bm p_N,t)} = \| \mathbf Z(\bm r_N) \|.
\label{metric}
\end{equation}
Statistical averages are therefore defined for the non\mbox-Hamiltonian
system as\cite{Fixman,BlueMoon}
\begin{eqnarray}
  \langle X (\bm r_N,\bm p_N) \rangle 
  &=& \frac{1}{Q} \int d\bm p_N
  d\bm r_N \, \|
  \mathbf Z(\bm r_N) \| \nonumber
\\
  &&\times\,
  X(\bm r_N, \bm p_N) \,
  \rho(\bm r_N,\bm p_N) \nonumber 
\\
  &&\times \prod_\alpha
  \delta( \sigma_\alpha (\bm r_N)) \delta \bigl(
  \dot{\sigma}_\alpha(\bm r_N,\bm p_N) \bigr),
\end{eqnarray}
where $\rho(\bm r_N,\bm p_N)$ is the probability density for the
unconstrained system and $Q$ is the partition function for the
constrained system, given by
\begin{eqnarray*}
  Q &=& \int d\bm p_N
  d\bm r_N \, \|
  \mathbf Z(\bm r_N) \| \, \rho(\bm r_N,\bm p_N)
  \nonumber
\\
  &&\times
  \prod_\alpha
  \delta( \sigma_\alpha (\bm r_N)) \delta \left(
  \dot{\sigma}_\alpha(\bm r_N,\bm p_N) \right).
\end{eqnarray*}
Although the invariant metric is non\mbox-uniform for many constrained
systems, for entirely rigid systems the $\mathbf Z(\bm r_N)$ matrix is
a function only of the point masses and fixed distances.  Hence the
term $\| \mathbf Z(\bm r_N) \|$ acts as a multiplicative factor which
cancels in the averaging process.

\subsection{Free motion of rigid bodies}
\label{without}

Although the solution of the dynamics of constrained systems via
time-independent holonomic constraints is intellectually appealing and
useful in developing a formal statistical mechanics for these systems,
it is often difficult to analytically solve for the values of the
Lagrange multipliers at arbitrary times.  One therefore often resorts
to numerical solutions of the multipliers in iterative form, using
algorithms such as SHAKE\cite{Ryckaert}.  Such an approach is not
really consistent with the principles of DMD, in which a
computationally efficient means of calculating event times is one of
the great advantages of the method.  For fully\mbox-constrained, rigid
bodies, it is more sensible to apply other, equivalent, approaches,
such as the principal axis or quaternion methods, to calculate
analytically the evolution of the system in the absence of external
forces.

The basic simplification in the dynamics of rigid bodies results from
the fact that the general motion of a rigid body can be decomposed
into a translation of the center of mass of the body plus a rotation
about the center of mass. The orientation of the body relative to its
center of mass is described by the relation between the so\mbox-called
\emph{body frame}, in which a set of axes are held fixed with the body
as it moves, and the fixed external \emph{laboratory frame}.  The two
frames of reference can be connected by an orthogonal transformation,
such that the position of an atom $i$ in a rigid body can be written
at an arbitrary time $t$ as:
\begin{equation}
  \bm r_i(t) = \bm R(t) + \mathbf A^\dagger(t) \, \tilde{\bm r}_i,
\label{cartesianpositions}
\end{equation}
where $\tilde{\bm r}_i$ is the position of atom $i$ in the body frame
(which is independent of time), $\bm R(t)$ is the center of mass, and
the matrix $\mathbf A^\dagger(t)$ is the orthogonal matrix that
converts coordinates in the body frame to the laboratory frame. Note
that matrix\mbox-vector and matrix\mbox-matrix multiplication will be implied
throughout the paper. The matrix $\mathbf A^\dagger(t)$ is the
transpose of $\mathbf A(t)$, which converts coordinates from the
laboratory frame to the body frame at time $t$.  The elements
composing the columns of the matrix $\mathbf A^\dagger(t)$ are simply
the coordinates of the axes in the body frame written in the
laboratory frame.  Note that Eq.~(\ref{cartesianpositions}) implies
that the relative vector $\overline{\bm r}_i(t)$ satisfies
\begin{equation}
  \overline{\bm r}_i =
  \bm r_i - \bm R
  = \mathbf A^\dagger \, \tilde{\bm r}_i,  
\label{relcartesianpositions}
\end{equation}
Here as well as below, we have dropped the explicit time dependence
for most time dependent quantities with the exception of quantities at
time zero or at a time that is integrated over.

One sees that in order to determine the location of different parts of
the body in the laboratory frame, the rotation matrix $\mathbf A$ must
be specified. This matrix satisfies a differential equation that will
now be derived and subsequently solved.

Before doing so, it will be useful to restate some properties of
rotation matrices and establish some notation to be used
below. Formally, a rotation matrix $\mathbf U$ is an orthogonal matrix
with determinant one and whose its inverse is equal to its transpose
$\mathbf U^\dagger$. Any rotation can be specified by a rotation axis
$\hat{\bm n}=(n_1,n_2,n_3)$ and an angle $\theta$ over which to
rotate. Here $\hat{\bm n}$ is a unit vector, so that one may also say
that any non\mbox-unit vector $\theta\hat{\bm n}$ can be used to specify a
rotation, where its norm is equal to the angle $\theta$ and its
direction is equal to the axis $\hat{\bm n}$. According to Rodrigues'
formula, the matrix corresponding to this rotation is\cite{Goldstein}
\begin{widetext}
\begin{equation}
  \mathbf U(\theta\hat{\bm n}) 
  =
  \begin{pmatrix}
    n_1^{2} + \left( n_2^2+n_3^2 \right) \cos\theta &
    n_1 n_2 \left( 1 - \cos\theta \right) - n_3\sin\theta &
    n_1 n_3 \left( 1 - \cos\theta \right) + n_2\sin\theta \\
    n_1 n_2 \left( 1 - \cos\theta \right) + n_3\sin\theta &
    n_2^{2} + \left( n_1^2+n_3^2 \right) \cos\theta &
    n_2 n_3 \left( 1 - \cos\theta \right) - n_1 \sin\theta \\
    n_3 n_1 \left( 1 - \cos\theta \right) - n_2 \sin\theta &
    n_3 n_2 \left( 1 - \cos\theta \right) + n_1 \sin\theta &
    n_3^{2} + \left( n_1^2+n_2^2 \right) \cos\theta 
  \end{pmatrix}.
\label{generalRotation}
\end{equation}
\end{widetext}

The derivation of the differential equation for $\mathbf A$ starts by
taking the time derivative of Eq.~\eqref{cartesianpositions}, yielding
\begin{equation}
  \bm v_i - \bm V
  = \dot{\mathbf A}^\dagger\,\tilde{\bm r}_i
  = \dot{\mathbf A}^\dagger\,\mathbf A\, \overline{\bm r}_i.
\label{step1vel}
\end{equation}
{}From elementary classical mechanics\cite{Goldstein}, it is known that
this relative velocity can also be written as
\begin{equation}
  \bm v_i - \bm V = \bm\omega \times \overline{\bm r}_i,
\label{step2vel}
\end{equation}
where $\bm\omega$ is the angular velocity vector in the lab frame.
Since both Eq.~\eqref{step1vel} and Eq.~\eqref{step2vel} are true for
any vector $\overline{\bm r}_i$, it follows that $\dot{\mathbf
A}^\dagger\,\mathbf A$ is the matrix representation of a cross product
with the angular velocity $\bm\omega$, i.e.,
\begin{equation}
  \dot{\mathbf A}^\dagger\,\mathbf A = \begin{pmatrix}
    0   & -\omega_3 &  \omega_2 \\
    \omega_3 & 0    & -\omega_1 \\
    -\omega_2 & \omega_1  & 0
  \end{pmatrix}
  \equiv \mathbf W(\bm\omega).
\label{step2b}
\end{equation}
Multiplying Eq.~\eqref{step2b} on the right with $\mathbf A^\dagger$
and taking the transpose on both sides (note that $\mathbf W$ is
antisymmetric) yields
\begin{equation}
   \dot{\mathbf A} = -
   \mathbf A\, \mathbf W(\bm\omega) .
\label{eom0}
\end{equation}
This equation involves the angular velocity in the laboratory frame,
but the rotational equations of motion are more easily solved in the
body frame.  The angular velocity vector transforms to the body frame
according to
\begin{equation}
  \tilde{\bm\omega} = \mathbf A\, \bm\omega.
\label{labAngVel}
\end{equation}
For any rotation $\mathbf A$ and vector $\bm x$ one has $\mathbf
W(\mathbf A\,\bm x)=\mathbf A\,\mathbf W(\bm x)\,\mathbf A^\dagger$,
hence one can write
\begin{equation}
  \mathbf W(\bm\omega)=
  \mathbf W(\mathbf A^\dagger\, \tilde{\bm\omega})
  =
  \mathbf A^\dagger\,\mathbf W(\tilde{\bm\omega})\,\mathbf A.
\label{Wlabproperty}
\end{equation}
Substituting Eq.~\eqref{Wlabproperty} into Eq.~\eqref{eom0} yields the
differential equation for $\mathbf A$:
\begin{equation}
  \dot{\mathbf A} = -\mathbf W(\tilde{\bm\omega})\, \mathbf A. 
\label{transformEquation}
\end{equation}

Although the choice of body frame is arbitrary, perhaps the most
convenient choice of axes for the body are the so\mbox-called principal
axes in which the moment of inertia tensor $\tilde{\mathbf I}$ is
diagonal, i.e., $\tilde{\mathbf I}=\mathrm{diag}(I_1,I_2,I_3)$.
Choosing this reference frame as the body frame, the representation of
the components of the angular momentum $\tilde{\bm L}$ is
\begin{equation}
  \tilde{\bm L} = \tilde{\mathbf I} \, \tilde{\bm\omega} 
  = \begin{pmatrix}
    I_1 \tilde{\omega}_1\\I_2 \tilde{\omega}_2\\
    I_3 \tilde{\omega}_3  \end{pmatrix},
\label{angularMomentum}  
\end{equation}
where $I_k$ and $\tilde{\omega}_k$ are the principal moments of
inertia and principal components of the angular velocity.

The time dependence of the principal components of the angular
velocity may be obtained from the standard expression for the torque
in the laboratory frame:
\begin{eqnarray}
  \bm\tau = \dot{\bm L} &=&
  \dot{\mathbf A}^\dagger \,\tilde{\bm L}  
  + \mathbf A^\dagger \,\dot{\tilde{\bm L}}  \nonumber 
\\
  &=&\mathbf A^\dagger \,\left[\tilde{\bm\omega} \times\tilde{\bm L}  
  + \dot{\tilde{\bm L}} \right].
\label{transformangularmomentum}
\end{eqnarray}
where Eq.~\eqref{transformEquation} was used in the last equality.
Transforming Eq.~\eqref{transformangularmomentum} to the principal
axis frame gives Euler's equations of motion for a rigid body
\begin{eqnarray}
  I_1 \dot{\tilde{\omega}}_1 - \tilde{\omega}_2\tilde{\omega}_3 (I_2-I_3) 
  &=& \tilde{\tau}_1 \nonumber 
\\
  I_2 \dot{\tilde{\omega}}_2 - \tilde{\omega}_1\tilde{\omega}_3 (I_3-I_1) 
  &=& \tilde{\tau}_2\label{Euler}  
\\
  I_3 \dot{\tilde{\omega}}_3 - \tilde{\omega}_1\tilde{\omega}_2 (I_1-I_2) 
  &=& \tilde{\tau}_3 ,\nonumber
\end{eqnarray}
where $\tilde{\tau}_k$ are the components of the torque
$\tilde{\bm\tau}=\mathbf A\,\bm\tau$ in the body frame.  Note that
even in the absence of any torque, the principal components of the
angular velocity are in general time dependent.

Once the angular velocity $\tilde{\bm\omega}$ is known, it can be
substituted into Eq.~\eqref{transformEquation} for the matrix $\mathbf
A$. The general solution of Eq.~\eqref{transformEquation} is of the
form
\begin{equation}
  \mathbf  A = \mathbf P\, \mathbf A(0).
\label{Asolution}
\end{equation}
where $\mathbf P$ is a rotation matrix itself which `propagates' the
orientation $\mathbf A(0)$ to the orientation at time $t$. $\mathbf P$
satisfies the same equation \eqref{transformEquation} as $\mathbf A$,
but with initial condition $\mathbf P(0)=\mathbbm1$.  By integrating
this equation, one can obtain an expression for $\mathbf P$.  At first
glance, it may seem that $\mathbf P$ can only be written as a formal
expression containing a time\mbox-ordered exponential.  However, for the
torque\mbox-free case $\bm\tau=0$, the conservation of angular
momentum and energy and the orthogonality of the matrix $\mathbf P$
can be used to
derive the following explicit expression\cite{VanZonSchofieldtoappear}
(implicitly also found in Ref.~\onlinecite{Masutanietal94}):
\begin{equation}
  \mathbf P = \mathbf T_1 \, \mathbf T_2.
\label{Pmat}
\end{equation}
Here $\mathbf T_1$ and $\mathbf T_2$ are two rotation matrices.  The
matrix $\mathbf T_1$ rotates $\tilde{\bm L}(0)$ to $\tilde{\bm L}$ and
can be written as
\begin{equation}
\mathbf T_1 = 
\begin{pmatrix}
  c_1c_2-s_1s_2c_3  & c_1s_2+s_1c_2c_3  & s_1s_3 \\
  -s_1c_2-c_1s_2c_3 & -s_1s_2+c_1c_2c_3 & c_1s_3 \\
  s_2s_3            & -c_2s_3           & c_3
\end{pmatrix}
 ,
\label{Q1}
\end{equation}
where
\begin{align}
s_1 &= \frac{\tilde L_1}{\tilde L_\perp}  
&
c_1 &= \frac{\tilde L_2}{\tilde L_\perp}    
\\
s_2 &= -\frac{\tilde L_1(0)}{\tilde L_\perp(0)}  
&
c_2 &= \frac{\tilde L_2(0)}{\tilde L_\perp(0)}  
\\
s_3 &= \frac{\tilde L_\perp\tilde L_3(0)-\tilde L_3\tilde
  L_\perp(0)}{L^2}
&
c_3 &= \frac{\tilde L_\perp\tilde L_\perp(0)+\tilde L_3\tilde
  L_3(0)}{L^2}
\end{align}
and $\tilde{L}_{\perp}=\sqrt{\tilde{L}_1^2+\tilde{L}_2^2}$ while $L
=|\tilde{\bm L}|$.

The matrix $\mathbf T_2$ can be expressed, using the notation in
Eq.~\eqref{generalRotation}, as
\begin{align}
  \mathbf T_2 &=\mathbf U(-\psi L^{-1} {\tilde{\bm L}}(0)) ,
\label{Q2}
\end{align}
where the angle $\psi$ is given by
\begin{equation}
  \psi = \int_0^t\! dt' \: \Omega(t') ,
\end{equation}
with
\begin{equation}
  \Omega=
  L\frac{I_1\tilde\omega_1^2+I_2\tilde\omega_2^2}{\tilde L_\perp^2}.
\label{Omegab}
\end{equation}
The angle $\psi$ can be interpreted as an angle over which the
body rotates. If the body rotates one way, the laboratory frame as
seen from the body frame rotates in the opposite way, which explains
the minus sign in Eq.~\eqref{Q2}.  For the derivation of
Eqs.~\eqref{Pmat}\mbox-\eqref{Omegab} we refer to
Ref.~\onlinecite{VanZonSchofieldtoappear}. Similar equations, but in a
special reference frame, can be found in
Ref.~\onlinecite{Masutanietal94}.

In the following, the solution of Eq.~\eqref{Euler} with $\bm\tau=0$
for bodies of differing degrees of symmetry will be analyzed and then
used to obtain explicit expressions for the matrix $\mathbf P$ as a
function of time and of the initial angular velocity in the body frame
$\tilde{\bm\omega}(0)$.

\subsubsection*{Case 1. Spherical rotor}

For the case of a spherical rotor in which all three moments of
inertia are equal, $I_1=I_2=I_3$, the form of the Euler equations
\eqref{Euler} is particularly simple:
$I_1\dot{\tilde{\omega}}_j=0$. It is therefore clear that all
components of the angular velocity in the body frame are conserved, as
are those of the angular momentum. As a result, $\mathbf T_1$ in
Eq.~\eqref{Q1} is equal to the identity matrix. A second consequence
is that $\Omega$ in Eq.~\eqref{Omegab} is constant, so that $ \psi =
\Omega\, t$ where $\Omega$ may be rewritten, using $I_1=I_2=I_3$, as $
\Omega = {L}/{I_1} = |\bm\omega|$.  Therefore Eqs.~\eqref{Pmat} and
\eqref{Q2} give
\begin{align}
  \mathbf P &=
  \mathbf U(-\bm \omega t),
                                 \label{Pmatrix}
\end{align}
corresponding to a rotation by an angle of $-\Omega t$ around the axis
$\bm\omega/\Omega$.

\subsubsection*{Case 2. Symmetric top}

For the case of a symmetric top for which $I_1=I_2$, one can solve the
Euler equations \eqref{Euler} in terms of simple sines and cosines,
since Eq.~\eqref{Euler} becomes
\begin{equation}
\begin{split}
  \dot{\tilde{\omega}}_1 &= \omega_p\,\tilde{\omega}_2
  \\
  \dot{\tilde{\omega}}_2 &= -\omega_p\,\tilde{\omega}_1
  \\
  \dot{\tilde{\omega}}_3 &= 0,
\end{split}
\label{symtop}
\end{equation}
where $\omega_p = \Big(1-\frac{I_3}{I_1}\Big)\tilde{\omega}_{3}(0)$ is
the precession frequency.  The full solution of the Euler
equations~\eqref{symtop} is given by
\begin{equation}
\label{symomegat}
 \tilde{\bm\omega} = 
\begin{pmatrix}
\tilde{\omega}_{1}(0)\cos \omega_pt 
  +\tilde{\omega}_{2}(0)\sin\omega_p t                        
\\
 -\tilde{\omega}_{1}(0)\sin \omega_pt 
  +\tilde{\omega}_{2}(0)\cos\omega_p t                        
\\
 \tilde{\omega}_{3}(0)
\end{pmatrix}.
\end{equation}
Using Eq.~\eqref{symomegat} and the fact that $\tilde{L}_\perp$ and
$\tilde L_3$ are conserved in this case, one can easily show that
$\mathbf T_1$ is given by
\begin{equation}
  \mathbf T_1 = 
  \begin{pmatrix}
  \cos\omega_p t &\sin\omega_p t &0\\
  -\sin\omega_p t&\cos\omega_p t&0\\
  0&0&1
  \end{pmatrix}
=\mathbf U(-\omega_pt\hat{\bm z}).
\end{equation}
and one can determine $\Omega$ from Eq.~\eqref{Omegab}:
\begin{equation}
  \Omega = 
  \frac{L[I_1{\tilde{\omega}_{1}}^2(0)
  +I_1{\tilde{\omega}_{2}}^2(0)]}
  {I_1^2{\tilde{\omega}_{1}}^2(0)
  +I_1^2{\tilde{\omega}_{2}}^2(0)}
  = \frac{L}{I_1}.
\end{equation}
This is a constant so that $ \psi = \frac{L}{I_1}t$.  Thus
\begin{equation}
  \mathbf T_2 = \mathbf U\left(-\frac{\tilde{\bm L}(0) t}{I_1}\right),
\end{equation}
and one gets from Eq.~\eqref{Pmat}:
\begin{equation}
  \mathbf P = 
  \mathbf U(-\omega_p t\hat{\bm z}) \, \mathbf
  U\left(-\frac{\tilde{\bm L}(0) t}{I_1}\right).
\end{equation}

\subsubsection*{Case 3. Asymmetric body}

If all the principal moments of inertia are distinct, the time
dependence of the angular velocity $\tilde{\bm\omega}$ involves
elliptic functions\cite{Goldstein}. While this may seem complicated,
efficient standard numerical routines exist to evaluate these
functions\cite{WhittakerWatson,AbramowitzStegun,NumRecipes,Moshier89,gsl}.
More challenging is the evaluation of the matrix $\mathbf P$.  While
its exact solution has been known for more than 170
years\cite{Rueb1834,Jacobi1850}, it is formulated\mbox--even in more
recent texts\cite{Whittaker,LandauLifshitzM}\mbox--in terms of
undetermined constants and using complex algebra, which hinders its
straightforward implementation in a numerical simulation.  It is
surprisingly difficult to find an explicit formula in the literature
for the matrix $\mathbf P$ as a function of the initial conditions,
which is the form needed in DMD simulations. For this reason, the
explicit general solution for $\mathbf P$ will briefly be presented
here in terms of general initial conditions. The details of the
derivation can be found elsewhere\cite{VanZonSchofieldtoappear}.

Following Jacobi\cite{Jacobi1850}, it is useful to adopt the
convention that $I_2$ is the moment of inertia intermediate in
magnitude (i.e., either $I_1<I_2<I_3$ or $I_1>I_2>I_3$) and one
chooses the overall ordering of magnitudes, such that:
\begin{equation}
\label{convention}
\begin{split}
  I_1>I_2>I_3  &\ \mbox{ if } E_R>\frac{L^2}{2I_2}\\
  I_1<I_2<I_3  &\ \mbox{ if } E_R<\frac{L^2}{2I_2} ,
\end{split}
\end{equation}
where $E_R$ is the rotational kinetic energy
$E_R=\frac12(I_1\tilde{\omega}_1^2+\tilde{\omega}_2^2+I_3\tilde{\omega}_3^2)$
and $L$ is the norm of the angular momentum $L=(I_1^2\tilde{\omega}_1^2 +
I_2^2\tilde{\omega}_2^2 + I_3^2\tilde{\omega}_3^2)^{1/2}$.  Without
this convention some quantities defined below would be complex valued,
which is numerically inconvenient and inefficient.  Note that in a
simulation molecules will often be assigned a specific set of physical
inertial moments with fixed order, i.e. not depending on the
particular values of $E_R$ and $L$. A simple way to nevertheless adopt
the convention in Eq.~\eqref{convention} is to introduce internal
variables $\tilde{\bm\omega}_{int}=\mathbf T_{int}\tilde{\bm\omega}$,
$\tilde{\mathbf I}_{int}=\mathbf T_{int}\tilde{\mathbf I}\mathbf
T_{int}$ and $\mathbf A_{int}=\mathbf T_{int}\tilde{\mathbf A}\mathbf
T_{int}$ which differ when necessary from the physical ones by a
rotation given by the rotation matrix
\begin{equation}
  \mathbf T_{int} =\begin{pmatrix}0&0&1\\0&-1&0\\1&0&0\end{pmatrix}.
\end{equation}
This matrix interchanges the $x$ and $z$ directions and reversed the
$y$ direction, and is equal to its inverse.

The Euler equations \eqref{Euler} can be solved because there are two
conserved quantities $E$ and $L^2$ which allow $\tilde{\omega}_1$ and
$\tilde{\omega}_3$ to be expressed in terms of $\tilde{\omega}_2$, at
least up to a sign which the quadratic conserved quantities cannot
prescribe. In this way the three coupled equations \eqref{Euler} are
reduced to a single ordinary differential equation for
$d\tilde{\omega}_2/dt$, from which $t$ can be solved as an integral
over $\tilde{\omega}_2$: This is an incomplete elliptic integral of
the first kind\cite{WhittakerWatson,AbramowitzStegun}. To get
$\tilde{\omega}_2$ as a function of $t$, one needs its inverse, which
is the elliptic function $\sn$\cite{WhittakerWatson,AbramowitzStegun}.
Without giving further details, the solution of the Euler equations is
given
by\cite{Jacobi1850,Whittaker,MarsdenRatiu,VanZonSchofieldtoappear}
\begin{equation}
\tilde{\bm\omega} = 
\begin{pmatrix}
\omega_{1m} \cn(\omega_p t+\varepsilon|m)
\\
 \omega_{2m} \sn(\omega_p t+\varepsilon|m)
\\
 \omega_{3m} \dn(\omega_p t+\varepsilon|m)
\end{pmatrix}.
\label{bodyAngvel}
\end{equation}
Here $\cn$ and $\dn$ are also elliptic
functions\cite{Knopp2,WhittakerWatson,AbramowitzStegun}, while the
$\omega_{im}$ are the extreme (maximum or minimum) values of the
$\omega_i$ and are given by
\begin{equation}
\begin{split}
  \omega_{1m} &= \sgn(\tilde{\omega}_{1}(0))\sqrt{\frac{L^2-2I_3E_R}{I_1(I_1-I_3)}}
\\
  \omega_{2m} &= -\sgn(\tilde{\omega}_{1}(0))\sqrt{\frac{L^2-2I_3E_R}{I_2(I_2-I_3)}}
\\
  \omega_{3m} &= \sgn(\tilde{\omega}_{3}(0))\sqrt{\frac{L^2-2I_1E_R}{I_3(I_3-I_1)}},
\end{split}
\end{equation}
where $\sgn(x)$ is the sign of $x$. Furthermore, in
Eq.~\eqref{bodyAngvel} the \emph{precession frequency} $\omega_p$ is
given by
\begin{equation}
  \omega_p  =
  \sgn(I_2-I_3)  \sgn(\tilde{\omega}_{3}(0))\sqrt{\frac{(L^2-2I_1E_R)(I_3-I_2)}{I_1I_2I_3}}.
\label{genomegap}
\end{equation}
The elliptic functions are periodic functions of their first argument,
and look very similar to the sine, cosine and constant function. They
furthermore depend on the \emph{elliptic parameter} $m$ (or elliptic
modulus $\sqrt{m}$), which determines how closely the elliptic
functions resemble their trigonometric counterparts, and which is
given by
\begin{equation}
  m= \frac{(I_1-I_2)(L^2-2I_3E_R)}{(I_3-I_2)(L^2-2I_1E_R)}.
\label{modulus}
\end{equation}

\begin{figure}[t]
\includegraphics[width=0.85\columnwidth]{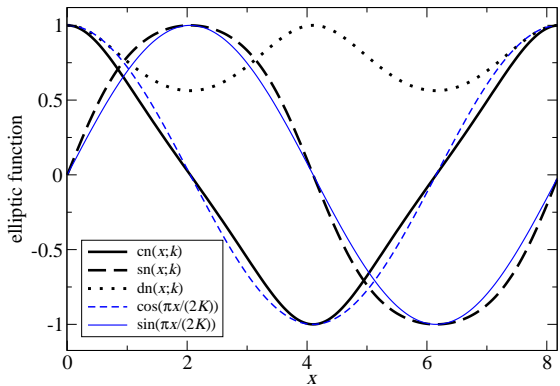}
\caption{Example of the elliptic functions $\cn$ (solid line),
  $\sn$ (bold dashed line), $\dn$ (dotted line)
  for $m=0.465$ ($K=2.05$, $K'=1.72$, $q=0.071$). Also plotted are the
  cosine (short dashed line)  and sine (thin short dashed line) with
  the same period, for comparison.}
\label{cnsndn}
\end{figure}

By matching the values of $\tilde{\omega}_2$ at time zero, one can
determine the integration constant $\varepsilon$:
\begin{equation}
\varepsilon = F\bigl(\tilde{\omega}_{20}/\tilde{\omega}_{2m}|m\bigr),
\end{equation}
where $F$ is the incomplete elliptic integral of the first
kind\cite{WhittakerWatson,AbramowitzStegun}
\begin{equation}
  F(y|m) = \int_0^y \frac{dx}{\sqrt{(1-x^2)(1-mx^2)}}.
\label{F}
\end{equation}
In fact, $\sn(x|m)$ is simply the inverse of this function. As a
result of the ordering convention in Eq.~\eqref{convention}, the
parameter $m$ in Eq.~\eqref{modulus} is guaranteed to be less than
one, which is required in order that $F(y|m)$ in Eq.~\eqref{F} not be
complex\mbox-valued.

Three more numbers can be derived from the elliptic parameter $m$ which
play an important role in the elliptic functions. These are the
\emph{quarter\mbox-period} $K = F(1|m)$, the \emph{complementary
quarter\mbox-period} $K'=F(1|1-m)$ and the \emph{nome}
$q=\exp(-\pi K'/K)$, which is the parameter in various series
expansions.

The period of the elliptic functions $\cn$ and $\sn$ is equal to $4K$,
while that of $\dn$ is $2K$.  These elliptic functions have the
following Fourier series\cite{WhittakerWatson,AbramowitzStegun}:
{\allowdisplaybreaks
\begin{align}
  \cn(x|m) &=  \frac{2\pi}{\sqrt mK}\sum_{n=0}^\infty
  \frac{q^{n+1/2}}{1+q^{2n+1}}\cos \frac{(2n+1)\pi x}{2K}
\label{cn}
\\
  \sn(x|m) &= \frac{2\pi}{\sqrt mK}\sum_{n=0}^\infty
  \frac{q^{n+1/2}}{1-q^{2n+1}}\sin \frac{(2n+1)\pi x}{2K}
\\
  \dn(x|m) &= \frac{\pi}{2K}+\frac{2\pi}{K}\sum_{n=1}^\infty
  \frac{q^n}{1-q^{2n}}\cos \frac{n\pi x}{K}.
\label{dn}
\end{align}
Note that the right\mbox-hand side of Eqs.~\eqref{cn}\mbox-\eqref{dn} depends
on $m$ through $K=F(1|m)$ and $q=\exp[-\pi F(1|1-m)/K]$. For $m=0$,
one gets $q=0$ and $\cn$, $\sn$ and $\dn$ reduce to $\cos$, $\sin$ and
$1$, respectively.  The constancy of $\dn(x|m=0)$} is reminiscent of
the conservation of $\tilde\omega_3$ in the case of the symmetric top,
and, indeed, for $I_1=I_2$, $m=0$ according to Eq.~\eqref{modulus}.

Typical values for $q$ are quite small, hence often the elliptic
function $\sn$, $\cn$ and $\dn$ resemble the $\cos$, $\sin$ and a
constant function with value one (as e.g.\ in Fig.~\ref{cnsndn}). For
small values of $q$, the series expressions for the elliptic functions
converge quickly (although this is not the best way to compute the
elliptic
functions\cite{WhittakerWatson,AbramowitzStegun,NumRecipes,Moshier89,gsl}).

Having given the solutions of the Euler equations, we now turn to the
solution of Eq.~\eqref{transformEquation} as given by
Eqs.~\eqref{Asolution}\mbox-\eqref{Omegab}.  The expression on the
right\mbox-hand side of Eq.~\eqref{Omegab} isnot a constant in this case
but involves elliptic functions. Despite this difficulty, the integral
can still be performed using some properties of elliptic functions,
with the result\cite{VanZonSchofieldtoappear}
\begin{equation}
  \psi(t) = A_1 + A_2 t - \phi(t).
\end{equation}
The constants $A_1$, $A_2$ and the periodic function $\phi(t)$ can be
expressed using the theta function
$H(u|m)$\cite{Jacobi1850,WhittakerWatson,AbramowitzStegun} as
\begin{align}
\label{tra}
  \phi(t) &= \arg H\big(\omega_p t+\varepsilon-\mathrm i\eta|m\big)
\\
  A_1 &= \phi(0) = \arg H(\varepsilon-\mathrm i\eta|m)
\\
  A_2 &= \frac{L}{I_1}+  \omega_p
  \frac{d\log H(\mathrm i\eta|m)}{d\eta} ,
\label{trg}
\end{align}
where we have used the definition
\begin{equation}
\eta =
\sgn(\tilde\omega_{30})K' - F\left(\frac{I_3\omega_{3m}}{L}\Big|1-m\right).
\label{eta}
\end{equation}

The equations \eqref{tra}\mbox-\eqref{trg} involve complex values which
are not convenient for numerical evaluation. Using the known series
expansions of the theta function $H$ and its logarithmic
derivative\cite{WhittakerWatson,AbramowitzStegun} in terms of the nome
$q$, these equations may be rewritten in a purely real form.  In fact,
one readily obtains the sine and cosine of $\psi$, which are all that
is needed in Eqs.~\eqref{generalRotation} and \eqref{Q2},
{\allowdisplaybreaks
\begin{equation}
\begin{split}
  \cos\psi(t) &= \frac{h_r(t)\cos(A_1+A_2t)+h_i(t)\sin(A_1+A_2t)}
	  {\sqrt{h_r^2(t)+h^2_i(t)}}
\\
  \sin\psi(t) &= \frac{h_r(t)\sin(A_1+A_2t)-h_i(t)\cos(A_1+A_2t)}
	  {\sqrt{h_r^2(t)+h^2_i(t)}}.
\end{split}
\end{equation}
with
\begin{align}
  h_r(t)&= \re H(\omega_p t+\varepsilon-\mathrm i\eta|m)
\nonumber\\*
  &= 2q^{1/4}\sum_{n=0}^\infty (-1)^n q^{n(n+1)}
   \cosh \frac{(2n+1)\pi \eta}{2K}
   \nonumber\\*&\qquad\qquad\qquad\times
   \sin \frac{(2n+1)\pi(\omega_p t+\varepsilon)}{2K}
\label{Hr}
\\
  h_i(t) &= \im H\big(\omega_p t+\varepsilon-\mathrm i\eta|m\big)
  \nonumber\\*
  &=
  -2q^{1/4}\sum_{n=0}^\infty (-1)^n q^{n(n+1)}
  \sinh \frac{(2n+1)\pi \eta}{2K}
  \nonumber\\*&\qquad\qquad\qquad\times
  \cos \frac{(2n+1)\pi(\omega_p t+\varepsilon)}{2K},
\label{Hi}
\end{align}
while} the constant $A_1$ is
\begin{equation}
  A_1 =\arctan[h_i(0)/h_r(0)] + n\pi,
\end{equation}
where $n=0$ if $h_r(0)>0$, $n=1$ if $h_r(0)<0$ and $h_i(0)>0$, and
$n=-1$ if $h_r(0)<0$ and $h_i(0)<0$.  Finally, the constant $A_2$ is
given by\cite{WhittakerWatson,AbramowitzStegun}
\begin{equation}
  A_2 = \frac{L}{I_1} + \frac{\pi\omega_p}{2K}\left[
    \frac{\xi+1}{\xi-1}-2\sum_{n=1}^\infty 
    \frac{q^{2n}(\xi^n-\xi^{-n})}{1-q^{2n}}
    \right],
\label{C2}
\end{equation}
where $\xi=\exp(\pi \eta/K)$.  The series expansion in $q$ in
Eq.~\eqref{C2} convergences for $\xi q^2<1$. Because $-K'<\eta<K'$
(cf.~Eq.~\eqref{eta}), one has $\xi q^2\leq q<1$, and the series
always converges. Since $q$ is typically small, the convergence is
rarely very slow (e.g.\ for convergence up to relative order $\delta$
one needs $O(\log\delta/\log q)$ terms). Note that since the constants
$A_1$ and $A_2$ depend only on the initial angular velocities, they
only need to be calculated once at the beginning of the motion of a
free rigid body.  On the other hand, the series expansions in
Eqs.~\eqref{Hr} and \eqref{Hi}, which have to be evaluated any time
the positions are desired, have extremely fast convergence due to the
$q^{n(n+1)}$ appearing in these expressions (for example, unless
$m\gtrsim0.95$, the series converges up to $O(10^{-15})$ occurs taking
only three terms).

There are efficient routines to calculate the functions $\cn$, $\sn$
$\dn$ and $F$, see e.g.\ Refs.\
\onlinecite{WhittakerWatson,AbramowitzStegun,NumRecipes,Moshier89,gsl},
and the series in Eqs.~\eqref{Hr}, \eqref{Hi} and \eqref{C2} converge,
the former two quite rapidly in fact. Therefore, despite an apparent
preference in the literature for conventional numerical integration of
the equations of motion via many successive small time steps even for
torque\mbox-free cases, the analytical solution can be used to calculate
the same quantities in a computationally more efficient manner
requiring only the evaluation of special functions. The gain in
efficiency should be especially pronounced in applications in which
many evaluations at various times could be needed, such as in the root
searches in discontinuous molecular dynamics (see below).

\section{Detection of interaction events}
\label{IncludingCollisions}

\subsection{Calculation of interaction events}

If the interaction potential between atoms $i$ and
$j$ is assumed to be discontinuous, say of the form
\begin{equation}
 \Phi( | \bm r_i - \bm r_j | ) =
\begin{cases}
\Phi_0 & \text{if }| \bm r_i - \bm r_j | \leq d \\
\Phi_1 & \text{if }| \bm r_i - \bm r_j | > d, 
\end{cases}
\label{potential}
\end{equation}
then rigid molecules interacting via this potential evolve freely until there is a change in the
potential energy of the system and an \emph{interaction event} or
\emph{collision event} occurs.  The time at which an event
occurs is governed by a collision indicator function $f_{ij}=|\bm
r_j-\bm r_i|^2-d^2$ defined such that at time $t_c$,
$f_{ij}(t_c)=0$. Here, the time dependence of $\bm r_j$ and $\bm r_i$
can be obtained using the results of Sec.~\ref{Calculation}.

The simplest example of this kind of system consists of two hard
spheres of diameter $d$ located at positions $\bm r_j$ and $\bm
r_i$. If two spheres are approaching, when they get to a distance $d$
from one another, the potential energy would change from $\Phi_1$ to
$\Phi_0=\infty$ if they kept approaching one another. As this
eventually would lead to a violation of energy conservation, the
spheres bounce off one another in a \emph{hard\mbox-core collision} at
time $t_c$, where $t_c$ is determined by the criterion $f_{ij}(t_c)=|
\bm r_j(t_c) - \bm r_i(t_c)|^2 - d^2 = 0$, i.e. by the zeros of the
collision indicator function.  Another kind of interaction event, with
$\Phi_0$ and $\Phi_1$ finite, will be called a \emph{square\mbox-well
collision} because the potential then has a square well shape.

To find the times at which collisions take place, the zeros of the
collision indicator functions must be determined, which generally has
to be done numerically.  The calculation of the collision times of
non\mbox-penetrating rigid objects is an important aspect of manipulating
robotic bodies, and is also an important element of creating realistic
animation sequences.  As a result, many algorithms have been proposed
in these contexts to facilitate the event time
search\cite{Animation1,Animation2}.

The search for the earliest collision event time can be facilitated
using screening strategies to decide when rigid bodies may
overlap\cite{Animation1,Animation2}.  Usually, these involve placing
the bodies in bounding boxes and using an efficient method to
determine when bounding boxes intersect.  The simplest way to do this
in a simulation of rigid molecules is to place each molecule in the
smallest sphere around its center of mass containing all components of
the molecule\cite{CiccottiDiatomic}.  The position of the sphere is
determined by the motion of the center of mass, while any change in
orientation of the rigid molecule occurs within the sphere.
Collisions between rigid molecules can therefore only occur when their
encompassing spheres overlap, and the time at which this occurs can be
calculated analytically for any pair of molecules.  This time serves
as a useful point to begin a more detailed search for collision events
(see below).  Similarly, one can also calculate the time at which the
spheres no longer overlap, and use these event times to bracket a
possible root of the collision indicator function.  It is crucial to
make the time bracketing as tight as possible in any implementation of
DMD with numerical root searches because the length of the time
bracketing interval determines the required number of evaluations of
the positions and velocities of the atoms, and therefore plays a
significant role in the efficiency of the overall procedure.

The simplest reliable and reasonably efficient means of detecting a
root is to perform a \emph{grid search} that looks for changes in sign
of $f_{ij}$, i.e., one looks at $f_{ij}(t+n\Delta t)$ and
$f_{ij}(t+(n+1)\Delta t)$ for successive $n$. The time points
$t+n\Delta t$ will be called the \emph{grid points}. When a time
interval in between two grid points is found in which a sign change of
$f_{ij}$ occurs, the Newton\mbox-Raphson algorithm\cite{NumRecipes} can be
called to numerically determine the root with arbitrary accuracy.
Since the Newton\mbox-Raphson method requires the calculation of first
time derivatives, one must also calculate, for any time $t$, the
derivative $df_{ij}/dt=2\bm r_{ij} \cdot \bm v_{ij}$, where the
notation $\bm r_{ij}=\bm r_j-\bm r_i$ and $\bm v_{ij}=\bm v_j-\bm v_i$
has been used. Such time derivatives are readily evaluated using
Eqs.~(\ref{cartesianpositions}) and (\ref{step1vel}).

Unfortunately, while the Newton\mbox-Raphson method is a very efficient
algorithm for finding roots, it can be somewhat unstable when one is
searching for the roots of an oscillatory function. For translating
and rotating rigid molecules, the collision indicator function is
indeed oscillatory due to the periodic motion of the relative
orientation of two colliding bodies.  It is particularly easy to miss
so\mbox-called \emph{grazing collisions} when the grid search interval
$\Delta t$ is too large, in which case the indicator function is
positive in two consecutive points of the grid search, yet nonetheless
``dips'' below zero in the grid interval.  It is important that no
roots are missed, for a missed root can lead to a different, even
infinite energy (but see Sec.~\ref{Ensembles} below).  To reduce the
frequency of missing grazing collisions to zero, a vanishingly small
grid interval $\Delta t$ would be required. Of course such a scheme is
not practical, and one must balance the likelihood of missing events
with practical considerations since several collision indicator
functions need to be evaluated at each point of the grid.  Clearly the
efficiency of the root search algorithm significantly depends on the
magnitude of grid interval.

To save computation time, a coarser grid can be utilized if a means of
handling grazing collisions is implemented. Since the collision
indicator function has a local extremum (maximum or minimum, depending
on whether $|\bm r_i-\bm r_j|^2$ is initially smaller or larger than
$d^2$) at some time near the time of a grazing collision, a reasonable
strategy to find these kind of collision events is to determine the
extremum of the indicator function in cases in which the indicator
function $f_{ij}$ itself does not change sign on the interval but its
derivative $df_{ij}/dt$ does. Furthermore, since the indicator
function at the grid points near a grazing collision is typically
small, it is fruitful to search only for extrema when the indicator
function at one of the grid points lies below some threshold
value\cite{fnb}. To find the local extrema of the indicator function,
any simple routine of locating the extrema of a non\mbox-linear function
can be utilized.  For example, Brent's minimization
method\cite{NumRecipes, Brent}, which is based on a parabolic
interpolation of the function, is a good choice for sufficiently
smooth one\mbox-dimensional functions.  Once the extremum is found, it is
a simple matter to decide whether or not a real collision exists by
checking the sign of $f_{ij}$.
 
Once the root has been bracketed (either through a sign change of
$f_{ij}$ during the grid search or after searching for an extremum),
one can simply use the Newton\mbox-Raphson algorithm to find the root to
desired accuracy, typically within only a few iterations.  The time
value returned by the Newton\mbox-Raphson routine needs to be in the
bracketed interval and $df_{ij}/dt < 0$ if $f_{ij}$ was initially
positive and $df_{ij}/dt > 0$ if it was initially negative. If those
criteria are not satisfied, the Newton\mbox-Raphson algorithm has clearly
failed and a less efficient but more reliable method is needed to
track down the root. For example, the Van~Wijngaarden\mbox-Dekker\mbox-Brent
method\cite{NumRecipes, Brent}, which combines bisection and quadratic
interpolation, is guaranteed to converge if the function is known to
have a root in the interval under analysis.

\subsection{Scheduling events}

In the previous section is was shown how to determine the time
$t_{ij}$ at which two atoms $i$ and $j$ collide under the assumption
that there is no other earlier collision. This we will call a
\emph{possible collision event}.  In a DMD simulation, once the
possible collision events at times $t_{ij}$ have been computed for all
possible collision pairs $i$ and $j$, the earliest event
$t^*_{i^*j^*}=\min_{ij} t_{ij}$ should be selected. After the
collision event between atoms $i^*$ and $j^*$ has been executed
(according to the rules derived in the next section), the next
earliest collision should be performed. However, because the
velocities of atoms of the molecules involved in the collision have
changed, the previously computed collision times involving these
molecules are no longer valid. The next event in the sequence can be
determined and performed only after these collision times have been
recomputed.

This process describes the basic strategy of DMD, which without
further improvements would be needlessly inefficient.  For if $M$ is
the number of possible collision events, finding the earliest time
would require $O(M)$ checks, and $M=O(N^2)$, while the number of
invalidated collisions that have to be recomputed after each collision
would be $O(N)$. Since the number of collisions in the system per unit
of physical time also grows with $N$, the cost of a simulation for a
given physical time would be $O(N^2)$ for the computation of collision
times and $O(N^3)$ for finding the first collision event\cite{fnc}.
Fortunately, there are ways to significantly reduce this computational
cost\cite{Rapaport,AlderWainwright59,ErpenbeckWood77,Rapaport80,Lubachevsky91,Marinetal93,MarinCordero95}.
The first technique, also used in molecular dynamics simulations of
systems interacting with continuous potentials, reduces the number of
possible collision times that have to be computed by employing a
\emph{cell division} of the system\cite{AlderWainwright59}.  Note that
while the times of certain interaction events (e.g. involving only the
molecule's center of mass) can be expressed in analytical form and
thus computed very efficiently, the atom\mbox-atom interactions have, in
general, an orientational dependence and the possible collision time
has to be found by means of a numerical root search as explained in
the previous section.  As a consequence, the most time consuming task
in a DMD simulation with rigid bodies is the numerical root search for
the collision times.  One can however minimize the required number of
collision time computations by dividing the system into a cell
structure and sorting all molecules into these cells according to the
positions of their centers of mass. Each cell has a diameter of at
least the largest ``interaction diameter'' of a molecule as measured
from its center of mass. As a result, molecules can only collide if
they are in the same cell or in an adjacent cell, so the number of
collision events to determine and to recompute after a collision is
much smaller.  In this technique, the sorting of molecules into cells
is only done initially, while the sorting is dynamically updated by
introducing a \emph{cell\mbox-crossing event} for each molecule that is
also stored\cite{Rapaport,Rapaport80}.  Since the center of mass of a
molecule performs linear motion between collision events, one can
express its cell\mbox-crossing time analytically and therefore the
numerical computation of that time is very fast.

The second technique reduces the cost of finding the earliest event
time. It consists of storing possible collision and cell-crossing
events in a time\mbox-ordered structure called a \emph{binary tree}. For
details we refer to Refs.~\onlinecite{Rapaport} and
\onlinecite{Rapaport80} (alternative event scheduling algorithms
exist\cite{Lubachevsky91,Marinetal93} but it is not clear which
technique is generally the most efficient\cite{MarinCordero95}.)

Finally, a third standard technique is to update the molecules'
positions and velocities only at collisions (and possibly upon their
crossing the periodic boundaries), while storing the time of their
last collision as a property of the molecule called its \emph{local
clock}\cite{ErpenbeckWood77}. Whenever needed, the positions and
velocities at later times can be determined from the exact solution of
force\mbox-free and torque\mbox-free motion of the previous
Sec.~\ref{Calculation}.

The use of cell divisions, a binary event tree to manage the events,
and local clocks is a standard practice in DMD simulations and largely
improves the simulation's efficiency\cite{Rapaport}. To see this, note
that in each step of the simulation one picks the earliest event from
the tree, which scales as $O(\log N)$ for randomly balanced
trees\cite{Rapaport,Rapaport80}. If it is a cell\mbox-collision event, it
is then performed and subsequently $O(1)$ collisions and cell
crossings are recomputed and added to the event tree ($\propto O(\log
N)$). If it is a crossing event, the corresponding molecule is put in
its new cell, new possible collision and crossing events are computed
($(O(1)$) and added to the tree ($O(\log N)$). Then the program
progresses to the next event. Since still $O(N)$ real events take
place per unit of physical time, one sees that using these techniques,
the computational cost per unit of physical time due to the
computation of possible collisions and cell crossing times scales as
$O(N)$ instead of $O(N^2)$, while the cost due to the event scheduling
is $O(N\log N)$ per unit of physical time instead of $O(N^3)$ \mbox-- a
huge reduction.

Contrary to what their scaling may suggest, one often finds that the
cost of the computation of collision times greatly dominates the
scheduling cost for finite $N$.  This is due to fact that the
computations of many of the collision times requires numerical root
searches, although some can, and should, be done analytically. Thus,
to gain further computational improvements, one has to improve upon
the efficiency of the numerical search for collision event times.  A
non\mbox-standard time\mbox-saving technique that we have developed for this
purpose is to use \emph{virtual collision events}. In this case, the
grid search (see Sec.~\ref{IncludingCollisions}) for a possible
collision time of atoms $i$ and $j$ is carried out only over a fixed
small number of grid points, thus limiting the scope of the root
search to a small search interval. If no collisions are detected in
this search interval, a virtual collision event is scheduled in the
binary event tree, much as if it were a possible future collision at
the time of the last grid point that was investigated. If the point at
which the grid search is curtailed is rather far in the future, it is
likely this virtual event will not be executed because the atoms $i$
and $j$ probably will have collided with other atoms beforehand.  Thus,
computational work has been saved by stopping the grid search after a
few grid points. Every now and then however, atoms $i$ and $j$ will
not have collided with other atoms at the time at which the grid
search was stopped. In this case, the virtual collision event in the
tree is executed, which entails continuing the root search from the
point at which the search was previously truncated.  The continued
search again may not find a root in a finite number of grid points and
schedule another virtual collision, or it may now find a collision. In
either case the new event is scheduled in the tree.  This
virtual collision technique avoids the unnecessary computation of a
collision time that is so far in the future that it will not be
executed in all likelihood anyway, while at the same time ensuring
that if, despite the odds, that collision is to happen nonetheless, it
is indeed found and correctly executed. The trade\mbox-off of this
technique is that the event tree is substantially larger, slowing down
the event management. Due to the high cost of numerical root searches
however, the simulations presented in the accompanying paper showed
that using virtual collision events yields an increase in efficiency
between 25\% and 110\%, depending mainly on the system size.

\section{Collision rules}
\label{Rules}

At each moment of collision, the impulsive forces and torques lead to
discontinuous jumps in the momenta and angular momenta of the
colliding bodies.  In the presence of constraints, there are two
equivalent ways of deriving the rules governing these changes.  In the
first approach, the dynamics are treated by applying constraint
conditions to Cartesian positions and momenta.  This approach is
entirely general and is suited for both constrained rigid and
non\mbox-rigid motion.  In its generality, it is unnecessarily complicated
for purely rigid systems and is not suitable for continuum bodies.
The second approach, suitable for rigid bodies only, uses the fact
that only six degrees of freedom, describing the center of mass motion
and orientational dynamics are required to fully describe the dynamics
of an arbitrary rigid body.  The derivation therefore consists of
prescribing a collision process in terms of impulsive changes to the
velocity of the center of mass and impulsive changes to the angular
velocity.

\subsection{Constrained variable approach}

The general collision process in systems with discontinuous potentials
can be seen as a limit of the collision process for continuous systems
in which the interaction potential becomes infinitely steep.  A useful
starting point for deriving the collision rules is therefore to
consider the effect of a force applied to the overall change in the
momentum of any atom $k$:
\begin{eqnarray}
\bm p_k' &\equiv& \bm p_k(t_c + \Delta t) = \bm p_k +
\int_{t_c - \Delta t}^{t_c + \Delta t} \dot{\bm p}_k(t) dt \nonumber \\
&=& \bm p_k +
\int_{t_c - \Delta t}^{t_c + \Delta t} \bm F_k(\bm r_N(t)) dt,
\label{deltaP1}
\end{eqnarray} 
where $\bm F_k$ is the total force acting on atom $k$ and $\bm p_k
\equiv \bm p_k(t_c - \Delta t)$.  Furthermore, here and below the pre
and post\mbox-collision values of a quantity $\bm a$ are denoted by $\bm a$
and $\bm a'$, respectively.

For discontinuous systems, the intermolecular forces are impulsive and
occur only at an instantaneous collision time $t_c$. When atoms $i$
and $j$ collide, the interaction potential $\Phi$ depends only on the
scalar distance $r_{ij}$ between those atoms, so that the force on an
arbitrary atom $k$ is given by (without summation over $i$ and $j$)
\begin{eqnarray*}
- \deriv{\Phi(\bm r_N)}{\bm r_k} 
&=&
- \deriv{\Phi(\bm r_N)}{r_{ij}}
\deriv{r_{ij}}{\bm r_{ij}}
\cdot
\deriv{\bm r_{ij}}{\bm r_k}
\\
&=&
- \deriv{\Phi(\bm r_N)}{r_{ij}} 
\frac{1}{r_{ij}}
\bm r_{ij}(\delta_{jk}-\delta_{ik} ).
\end{eqnarray*}
Note that this is non\mbox-zero only for the atoms involved in the
collision, as expected.  Given that the force is impulsive, it may be
written as
\begin{equation}
- \deriv{\Phi(\bm r_N)}{\bm r_k} 
=
S\, \delta(t-t_c) \hat{\bm r}_{ij}(\delta_{jk}-\delta_{ik} ),
\end{equation}
where the scalar $S$ is the magnitude of the impulse (to be
determined) on atom $a$ in the collision.

In general, the constraint forces on the right\mbox-hand side of
Eq.~(\ref{ELeq}) must also have an impulsive component whenever
intermolecular forces are instantaneous in order to maintain the rigid
body constraints at all times.  We account for this by writing the
Lagrange multipliers as
\[
\lambda_\alpha = \nu_\alpha + \mu_\alpha\delta(t-t_c).
\]
Because $\lambda_\alpha$ enters into the equations of motion for all
atoms $k$ involved in the constraint $\sigma_\alpha$, there is an
effect of this impulsive constraint force on all those atoms.  Thus,
one can write for the force on a atom $k$ when atoms $i$ and $j$
collide:
\begin{multline}
\bm F_k(\bm r_N) =
-
\nu_\alpha \deriv{\sigma_\alpha(\bm r_N)}{\bm r_k}  
\\
\,+ \delta (t-t_c) \left[ S \hat{\bm r}_{ij} 
(\delta_{jk}-\delta_{ik})
- \mu_\alpha
\deriv{\sigma_\alpha(\bm r_N)}{\bm r_k} \right].
\label{forces}
\end{multline}
Substituting Eq.~(\ref{forces}) into~(\ref{deltaP1}), one finds that
 the term proportional to $\nu_\alpha$ vanishes in the limit that the
 time interval $\Delta t$ approaches zero, so that the post\mbox-collision
 momenta $\bm p_k'$ are related to the pre\mbox-collision momenta $\bm
 p_k$ by
\begin{equation}
\bm p_k' = \bm p_k - \mu_\alpha
\deriv{\sigma_\alpha(\bm r_N)}{\bm r_k}
 + S  \hat{\bm r}_{ij} (\delta_{jk}-\delta_{ik}).
\label{deltaP2}
\end{equation}
Note that at the instant of collision $t=t_c$, the positions of all
atoms $\bm r_N$ remain the same (only their momenta change) so that
there is no ambiguity in the right\mbox-hand side of Eq.~\eqref{deltaP2}
as to whether to take the $\bm r_N$ before or after the collision.  It
is straightforward to show that due to the symmetry of the interaction
potential, the total linear momentum and angular momentum of the
system are conserved by the collision rule~Eq.(\ref{deltaP2}) for
arbitrary values of the unknown scalar functions $S$ and $\mu_\alpha$.
In addition to these constants of the motion, the collision rule must
also conserve total energy and preserve the constraint conditions,
$\sigma_\alpha (\bm r_N) = 0$ and $\dot{\sigma}_\alpha(\bm r_N) = 0$,
before and after the collision.  The first constraint condition is
trivially satisfied at the collision time, since the positions are not
altered at the moment of contact.  The second constraint condition
allows the scalar $\mu_\alpha$ to be related to the value of $S$ using
Eq.~\eqref{mustobey} before and after the collision, since we must
have
\begin{eqnarray}
\dot{\sigma}_\alpha &=&
\sum_k \frac{\bm p_k}{m_k} \cdot
\deriv{\sigma_\alpha}{\bm r_k} = 0 \nonumber \\
\dot{\sigma}'_\alpha &=& \sum_k \frac{\bm p_k'}{m_k} \cdot
\deriv{\sigma_\alpha}{\bm r_k} = 0.
\label{40}
\end{eqnarray}
Inserting Eq.~(\ref{deltaP2}) into Eq.~\eqref{40}, one gets
\begin{eqnarray}
0 &=& \sum_k \frac{1}{m_k} \left( 
S \hat{\bm r}_{ij} (\delta_{jk}-\delta_{ik})
 -  \mu_\beta
\deriv{\sigma_\beta }{\bm r_k} \right) \cdot
\deriv{\sigma_\alpha}{\bm r_k}.
\end{eqnarray}
Solving this linear equation for $\mu_\alpha$ gives
\begin{equation}
\label{mus}
\begin{split}
\mu_\alpha &= \mathbf Z^{-1}_{\alpha \beta} \mathcal{F}_\beta 
\\
\mathcal{F}_\beta &=
S \hat{\bm r}_{ij}\cdot\left(
\frac{1}{m_j}\deriv{\sigma_\beta}{\bm r_j}
-
\frac{1}{m_i}\deriv{\sigma_\beta}{\bm r_i}
\right)
,
\end{split}
\end{equation}
where the $\mathbf Z$ matrix was defined in Eq.~(\ref{Zdef}). Note
that if atoms $i$ and $j$ are on different bodies, a given constraint
$\sigma_\beta$ involves either one or the other atom (or neither), so
at least one of the two terms on the right\mbox-hand side of
Eq.~\eqref{mus} is then zero.  Equation \eqref{deltaP2} can now be
written as
\begin{eqnarray}
\bm p_k' &=&
 \bm p_k + S \Delta\bm p_k
\nonumber\\
\Delta \bm p_k 
&=&
\hat{\bm r}_{ij}(\delta_{jk}-\delta_{ik})-\mu_\alpha^*
\deriv{\sigma_\alpha}{\bm r_k},
\label{changeP}
\end{eqnarray}
where $\mu_\alpha^* = \mu_\alpha/S$ is a function of the
phase\mbox-space coordinate as determined by Eq.~\eqref{mus} and is
independent of $S$.

Finally, the scalar $S$ can be determined by employing energy
conservation,
\begin{equation}
\frac{\bm p_k' \cdot \bm p_k' }{2m_k} + \Delta \Phi =
\frac{\bm p_k \cdot \bm p_k }{2m_k} ,
\label{thisone}
\end{equation}
where $\Delta \Phi = \Phi' - \Phi$ denotes the discontinuous
change in the potential energy at the collision time.  Inserting
the expression in~(\ref{changeP}) into \eqref{thisone} and using
Eq.~\eqref{mustobey}, one gets a quadratic equation for the scalar $S$,
\begin{eqnarray}
&&aS^2 + bS + \Delta \Phi = 0 
\nonumber\\
&&a = \sum_k\frac{\Delta \bm p_k\cdot \Delta \bm p_k}{2m_k}  \\
&&b = \sum_k \frac{\bm p_k \cdot \Delta \bm p_k}{m_k} =
 \hat{\bm r}_{ij} \cdot \bm v_{ij} \nonumber .
\label{quadratic}
\end{eqnarray}
For finite values
of $\Delta \Phi$, the value of $S$ is therefore
\begin{equation}
S = \frac{-b \pm \sqrt{b^2 - 4a\Delta \Phi}}{2a},
\end{equation}
where the physical solution corresponds to the positive (negative)
root if $b >0$ ($b<0$), provided $b^2 > 4a\Delta \Phi$.  If this
latter condition is not met, there is not enough kinetic energy to
overcome the discontinuous barrier, and the system experiences a
hard\mbox-core scattering, with $\Delta \Phi=0$, so that
Eq.~\eqref{quadratic} gives $S=-b/a$.  Once the value of $S$ has been
computed, the discrete changes in momenta or velocities are easily
computed using Eq.~(\ref{changeP}).

\subsection{Rigid body approach}

The solution method outlined above can be applied to semi\mbox-flexible as
well as rigid molecular systems, but is not very suitable for rigid,
continuous bodies composed of an infinite number of point particles.
For perfectly rigid molecules, a more convenient approach is therefore
to analyze the effect of impulsive collisions on the center of mass
and angular coordinates of the system, which are the minimum number of
degrees of freedom required to specify the dynamics of rigid systems.
The momentum of the center of mass $\bm P_a$ and the angular momentum
$\bm L_a$ of rigid molecule $a$ are affected by the impulsive
collision via
\begin{eqnarray}
\bm P_a' &=& \bm P_a + \Delta \bm P_a \nonumber \\
\bm L_a' &=& \bm R_a \times \bm P_a' +
\mathbf I_a \cdot \bm\omega_a'
\label{compare}
\\
&=&
\bm L_a + \bm R_a \times \Delta \bm P_a +\mathbf I_a
\cdot \Delta \bm\omega_a
\nonumber\\
&=&\bm L_a + \Delta\bm L_a ,
\nonumber
\end{eqnarray}
where $\mathbf I_a$ and $\bm\omega_a$ are the moment of inertia tensor
and the angular velocity of body $a$ in the laboratory frame,
respectively. Note that they are related to their respective
quantities in the principal axis frame (body frame) via the matrix
$\mathbf A_a(t)$ (now associated with the body~$a$):
\begin{eqnarray}
\mathbf I_a &=& \mathbf A_a^\dagger
\, \tilde{\mathbf I}_a \, \mathbf A_a
\nonumber \\
\bm\omega_a &=& \mathbf A_a^\dagger
\, \tilde{\bm\omega}_a.
\end{eqnarray}

To derive specific forms for the impulsive changes $\Delta \bm P_a$
and $\Delta\bm\omega_a$, one may either calculate the impulsive force
and torque acting on the center of mass and angular momentum, leading
to $\Delta\bm P_a=-\Delta\bm P_b=-S\hat{\bm r}$ and $\Delta\bm
L_a=-\Delta\bm L_b=\bm r_a\times\Delta\bm P_a$, where $\bm r_a$ and
$\bm r_b$ are the points at which the forces are applied on body $a$
and $b$, respectively, while $\hat{\bm r}=(\bm r_b-\bm r_a)/|\bm r_b-\bm
r_a|$, and $S$ should be obtained from energy conservation. To
understand this better and make a connection with the previous
section, one may equivalently view the continuum rigid body as a limit
of a non-continuum rigid body composed of $n$ constrained point
particles, and use the expressions derived in the previous section for
the changes in momenta of the constituents.

In the latter approach, it is convenient to switch the notation for
the positions and momenta of the atoms from $\bm r_i$ and $\bm p_i$ to
$\bm r^a_i$ and $\bm p^a_i$, which indicate the position and momentum
of particle $i$ on body $a$, respectively.  Using this notation and
considering a collision between particle $i$ on body $a$ and particle
$j$ on body $b$, Eq.~(\ref{changeP}) can be written as
\begin{eqnarray}
{\bm p_k^a}' - \bm p_k^a =S \Delta \bm p_k^a = 
-S \left[ \hat{\bm r}_{ij}^{ab}\delta_{ik} + \mu_\alpha^*
\deriv{\sigma_\alpha}{\bm r_k}\right],
\label{changePa}
\end{eqnarray}
where $\hat{\bm r}_{ij}^{ab}=\bm r^{ab}_{ij}/r^{ab}_{ij}$ is the unit
vector along the direction of the vector $\bm r^{ab}_{ij} = \bm
r_j^{b} - \bm r^a_i$ connecting atom $i$ on body $a$ with its
colliding partner $j$ on body $b$. Thus, noting that $\bm R_ a =
\sum_k m_k^a \bm r_k^a/M_a$, where $M_a = \sum_{k=1}^n m^a_k$ is the
total mass of body $a$, and using Eq.~(\ref{changePa}), one finds that
\begin{eqnarray}
\bm P_a' &=& \bm P_a + \Delta \bm P_a 
\label{changePCOM}\\
\Delta \bm P_a &=& 
S \sum_{k=1}^n \Delta \bm p_k^a = 
- S\mu^*_\alpha
\sum_{k=1}^n \deriv{\sigma_\alpha}{\bm r_k} 
-S\hat{\bm r}_{ij}^{ab} =  -S\hat{\bm r}_{ij}^{ab}  ,
\nonumber
\end{eqnarray}
since 
\begin{equation}
\sum_{k=1}^n \deriv{\sigma_\alpha (\bm r_{uv})}{\bm r_k} 
=
\sum_{k=1}^n \sigma_\alpha' \hat{\bm r}_{uv} \left( \delta_{ku} -
\delta_{kv} \right) = 0.
\end{equation}
Similarly, one finds that 
\begin{eqnarray}
\Delta \bm L_a &=& 
S \sum_{k=1}^n \bm r^a_k \times \Delta \bm p^a_k
\nonumber \\
&=& S \sum_{k=1}^n \left( \bm R_a + \overline{\bm r}^a_k \right) \times
\Delta \bm p_k^a \nonumber \\
&=& 
\bm R_a \times \Delta \bm P_a
+S \sum_{k=1}^n \overline{\bm r}^a_k \times \Delta
\bm p_k \nonumber \\
&=&  
\bm R_a \times \Delta \bm P_a -S \hspace{0.1cm}
 \overline{\bm r}^a_i \times
\hat{\bm r}^{ab}_{ij} ,
\end{eqnarray}
where $\overline{\bm r}^a_k = \bm r^a_k - \bm R_a$.  Comparing
with Eq.~\eqref{compare}, it is evident that
\begin{equation}
\Delta \bm\omega_a  = 
-S\, \mathbf I_a^{-1} \, \left(
\overline{\bm r}^a_i\times \hat{\bm r}^{ab}_{ij} \right) .
\label{changeOmega}  
\end{equation}
Note that $\mathbf I_a^{-1}$ is a matrix inverse.  Once again the
impulsive changes are directly proportional to $S$, and the change of
the angular velocity of body $b$ in the laboratory frame due to the
collision can be calculated analogously.

To determine the scalar $S$, one again uses the conservation of total
energy ($E'=E$) to see that
\begin{eqnarray}
&& \frac{|\bm P_a'|^2}{2M_a}
+ \frac{|\bm P_b'|^2}{2M_b} +
\frac{1}{2} \bm\omega_a' \cdot \mathbf I_a \, \bm\omega_a'
+
\frac{1}{2}\bm\omega_b' \cdot \mathbf I_b \, \bm\omega_b'
+ \Delta \Phi \nonumber \\
&&= 
\frac{|{\bm P}_a|^2}{2M_a}
+ \frac{|{\bm P}_b|^2}{2M_b} +
\frac{1}{2} \bm\omega_a \cdot \mathbf I_a \,\bm\omega_a 
+
\frac{1}{2}\bm\omega_b \cdot \mathbf I_b \, \bm\omega_b.
\end{eqnarray}
Inserting Eqs.~(\ref{changePCOM}) and (\ref{changeOmega}) into the
energy equation above yields, after some manipulation, a quadratic
equation for $S$ of the form of Eq.~(\ref{quadratic}), with
\begin{eqnarray}
a &=& \frac{1}{2M_a} + \frac{1}{2M_b} 
+ \frac{\Delta E^a_\omega + \Delta
E^{b}_\omega}{2} \nonumber \\
b 
&=&  \bm v^{ab}_{ij} \cdot \hat{\bm r}^{ab}_{ij},\nonumber
\end{eqnarray}
where
\begin{eqnarray*}
  \Delta E^a_\omega &=& \bm n_{ij}^a \cdot \mathbf I_a^{-1} 
  \, \bm n^a_{ij}
\\
  \Delta E^{b}_\omega &=& \bm n^{b}_{ij} \cdot \mathbf I_b^{-1} 
  \, \bm n^{b}_{ij},
\end{eqnarray*}
with 
\begin{eqnarray*}
  \bm n^a_{ij} &=& \overline{\bm r}_i^a \times \hat{\bm r}_{ij}^{ab}
\\
  \bm n^b_{ij} &=& \overline{\bm r}_j^{b} \times \hat{\bm r}_{ij}^{ab}.
\end{eqnarray*}
For a spherically symmetric system, $\mathbf I = \tilde{\mathbf I} =
I_1 \mathbbm1$, and $\Delta E^{a,b}_\omega = \bm n_{ij}^{a,b \dagger}
\cdot \bm n^{a,b}_{ij}/I_1$.

\section{Dynamics in the Canonical and Microcanonical Ensembles}
\label{Ensembles}

Any event\mbox-driven molecular dynamics simulation relies on the
assumption that no collision is ever missed.  However, collisions will
be missed whenever the time difference between two nearby events is on
the order of (or smaller than) the time error of the scheduled events,
which indicates that there is still a finite chance that a collision
is missed even when event times are calculated in a simulation
starting from an analytic expression, due to limits on machine
precision.  Although this subtle issue is not very important in a hard
sphere system, in the present context it is of interest.  Indeed, the
extensive use of numerical root searches for the event time
calculations combined with the need for computational efficiency
demands a lower precision in the time values of collision events
(typically a precision of $10^{-10}$ instead of $10^{-16}$ for
analytical roots).  In this section, it will be shown how to handle
missed collisions in the context of the hybrid Monte Carlo scheme
(HMC).

In general, the HMC method \cite{HMC1,HMC2} combines the Monte Carlo
method with molecular dynamics to construct a sequence of independent
configurations $\{ \bm r_N^{(1)},\dots,\bm r_N^{(n)} \}$, distributed
according to the canonical probability density
\begin{equation}
  \rho(\bm r_N) = \frac{1}{Z} \exp\left[ -\frac{\Phi(\bm
  r_N)}{k_\mathrm BT} \right],
\label{canonicalProb}
\end{equation}
where $Z$ is the configurational integral, $k_\mathrm B$ is
Boltzmann's constant, and $T$ is the temperature.  In the present
context, this method can be implemented as follows: Initially, a new
set of momenta $\bm p'_N$ is selected by choosing a random center of
mass momentum $\bm P$ and angular velocity $\bm\omega$ for each
molecule from the Gaussian distribution
\[
\rho_G \propto \exp \left[ -\frac{1}{k_\mathrm BT}\left(\frac{|\bm P|^2}{2M}+
\frac{1}{2}\bm\omega\cdot \mathbf I \, \bm\omega\right)\right].
\]
The system is then evolved deterministically through phase\mbox-space for
a fixed time $\tau_0$ according to the equations of motion.  This
evolution defines a mapping of phase\mbox-space given by $(\bm r_N(0),\bm
p_N(0)) \mapsto (\bm r_N(\tau_0),\bm p_N(\tau_0)) \equiv (\bm r'_N,\bm
p'_N)$.  The resulting phase space point $(\bm r'_N,\bm p'_N)$ and
trajectory segment are then accepted with probability
\begin{equation}
  p_A(\bm r'_N,\bm p'_N |\bm r_N(0),\bm p_N(0)) = \min \left\{ 1,
\exp\left[  -\frac{\Delta E}{k_\mathrm BT} \right]  
\right\} , 
\label{probAcceptance}
\end{equation}
where
\begin{equation}
\Delta E=E(\bm r'_N,\bm p'_N) - E(\bm r_N(0),\bm p_N(0)),
\label{deltaH}
\end{equation}
and
\begin{equation}
  E(\bm r_N,\bm  p_N) = \sum_{i=1}^N \frac{1}{2m_i}|\bm
  p_i|^2+\Phi(\bm r_N).
\end{equation}

This algorithm generates a Markov chain of configurations with an asymptotic
distribution given by the stationary canonical distribution defined in
Eq.~(\ref{canonicalProb}) provided that the phase space trajectory is
\emph{time reversible} and \emph{area preserving}\cite{HMC1}.  Since
free translational motion is time reversible, and the reversibility of
the rotational equations of motion is evident from Eq.~(\ref{Pmat}),
the first requirement is satisfied.  Furthermore, since the invariant
phase space metric is uniform for fully rigid bodies (see
Eq.~\eqref{metric} in Sec.~\ref{ConstrainedDynamics}), the \emph{area
preserving} condition is also satisfied.

Ideally, a DMD simulation satisfies $\Delta E=0$ so that according to
Eq.~(\ref{probAcceptance}) every trajectory segment is accepted. In
the less ideal, more realistic case in which collisions are
occasionally missed, the HMC scheme provides a rigorous way of
accepting or rejecting the segment. If a hard\mbox-core collision has been
missed and the configuration at the end of a trajectory segment has
molecules in unphysical regions of phase space where the potential
energy is infinite, then $\Delta E= \infty$ and the new configuration
and trajectory segment are always rejected.  On the other hand, if
only a square\mbox-well interaction has been missed, $\Delta E$ at the end
of the trajectory segment is finite and there is a non\mbox-zero
probability of accepting the trajectory.

An analogous strategy can be devised to carry out microcanonical
averages. In this case, the assignment of new initial velocities in
the first step is still done randomly but in such a way that the total
kinetic energy of the system remains constant. Such a procedure can be
carried out by exchanging center of mass velocities between randomly
chosen pairs of molecules.  The system is evolved dynamically through
phase space for a fixed time $\tau_0$ and the new phase space point
$(\bm r'_N,\bm p'_N)$ is accepted according to
\begin{equation}
p_A\left(\bm r'_N,\bm p'_N | \bm r_N(0),\bm p_N(0)\right) =
\begin{cases}
0 & \text{if } \Delta E \not= 0  \\
1 & \text{if } \Delta E = 0,
\end{cases}
\label{microcanonical}
\end{equation}
where $\Delta E$ is given by Eq.~(\ref{deltaH}). Clearly, the case
$\Delta E \not= 0$ only occurs when a collision has been missed, and
in such a case the trajectory segment is never accepted.

It should be emphasized that in the HMC scheme, a new starting
configuration for a segment of time evolution is chosen only after
every DMD time interval $\tau_0$.  An algorithm in which a new
configuration is selected only after a collision is missed is likely
to violate detailed balance, and is therefore not a valid Monte\mbox-Carlo
scheme.  On the other hand, the length of the trajectory segments
$\tau_0$ in the HMC method outlined above can be chosen to be slightly
larger than the relevant relaxation time of the system. Such a choice
allows one to use the deterministic phase space trajectory to compute
time\mbox-dependent correlation functions from the {\it exact} dynamics of
the system without rejecting a significant fraction of the trajectory
segments.

\section{Conclusions}
\label{Conclusions}

In this paper we have shown how to carry out discontinuous molecular
dynamics simulations for arbitrary semi\mbox-flexible and rigid
molecules. For semi\mbox-flexible bodies, the dynamics and collision rules
have been derived from the principles of constrained Lagrangian
mechanics.  The implementation of an efficient DMD method for
semi\mbox-flexible systems is hindered by the fact that in almost all
cases the equations of motion must be propagated numerically in an
event searching algorithm so that the constraints are enforced at all
times. Nonetheless, such a scheme can be realized using the
SHAKE\cite{Ryckaert} or RATTLE\cite{rattle} algorithms in combination
with the root searching methods outlined here.

The dynamics of a system of completely rigid molecules interacting
through discontinuous potentials is more straightforward.  For such a
system, the Euler equations for rigid body dynamics can be used to
calculate the free evolution of a general rigid object. This
analytical solution enables the design of efficient numerical
algorithms for the search for collision events. In addition, the
collision rules for calculating the discontinuous changes in the
components of the center of mass velocity and angular momenta have
been obtained for arbitrary bodies interacting through a point based
on conservation principles.  Furthermore, the sampling of the
canonical and microcanonical ensembles, as well as the handling of
missed collisions, has also been discussed in the context of a hybrid
Monte Carlo scheme.

{}From an operational standpoint, the difference between the method of
DMD and molecular dynamics using continuous potentials in rigid
systems lies in the fact that the DMD approach does not require the
calculation of forces and sequential updating of phase space
coordinates at discrete (and short) time intervals since the response
of the system to an impulse can be computed analytically. Instead, the
computational effort focuses on finding the precise time at which such
impulses exert their influence.  The basic building block outlined
here for the numerical computation of collision times is a grid
search, for which the positions of colliding atoms on a given pair of
molecules need to be computed at equally spaced points in time.  As
outlined in Sec.~\ref{IncludingCollisions}, this can be done
efficiently starting with a completely explicit analytical form of the
motion of a torque\mbox-free rigid body, without which the equations of
motions would have to be integrated numerically.  An efficient
implementation of the DMD technique to find the time collision events
should make use of a) a large grid step combined with a threshold
scenario to catch pathological cases, b) sophisticated but standard
techniques such as binary event trees, cell divisions, and local
clocks, and c) a new technique of finding collision times numerically
that involves truncating the grid search and scheduling virtual
collision events.

On a fundamental level, it is natural to wonder whether the `stepped'
form of a discontinuous potential could possibly model any realistic
interaction. Such concerns are essentially academic, since it is
always possible to approximate a given interaction potential with as
many (small) steps as one would like in order to approximate a given
potential to any desired level of accuracy\cite{Chapelaetal89}. Of
course, the drawback to mimicking a smooth potential with a
discontinuous one with many steps is that the number of `collision'
events that occur in the system per unit time scales with the number
of steps in the potential.  Hence, one would expect that the
efficiency of the simulation scales roughly inversely with the number
of steps in the interaction potential. Nonetheless, the issue is a
practical one: How small can the number of steps in the interaction
potential be such that one still gets a good description of the
physics under investigation? In the accompanying
paper\cite{following}, we will see for benzene and methane that it
takes surprisingly few steps (e.g.\ a hard core plus a square\mbox-well
interaction) to get results which are very close to those of
continuous molecular dynamics. Additionally, we compare the
efficiency of such simulations to simulations based on standard
molecular dynamics methods.

\section{Acknowledgments}

The authors would like to acknowledge support by grants from 
the Natural Sciences and Engineering Research Council of Canada
(NSERC).


\begin{thebibliography}{99}
\bibitem{fna} To be more precise, no Hamiltonian exists with $\bm
r_i$ and $\bm p_i$ as conjugate variables. Otherwise, $\dot{\bm r}_i$
should be equal to $\partial \mathcal H/\partial\bm p_i$ and $\dot{\bm
p}_i$ should be equal to $-\partial \mathcal H/\partial\bm r_i$,
whence $\partial\dot{\bm r}_i/\partial \bm r_i=-\partial\dot{\bm
p}_i/\partial\bm p_i$, but Eq.~\eqref{nheom} violates this relation.
\bibitem{fnb} The threshold value is found by trial and error; a
trial simulation with a very small threshold value is run which is
sure to miss a collision at some point due to a grazing collision; the
collision indicator function around the time of this grazing collision
is inspected and the threshold value is adjusted such that this
collision will not be missed. New trial simulation are run, and the
threshold adjusted, until the frequency of missed grazing collisions
is acceptable.
\bibitem{fnc} This assumes no optimization whatsoever. It is easy to
reduce the cost of finding the first possible collision times to
$O(N^2)$ per unit physical time, however, by storing only the first
collision for every given particle\cite{AllenTildesley}.
\end{thebibliography}
\end{document}